\documentclass[twocolumn,letterpaper,superscriptaddress,showpacs,prl]{revtex4}
\usepackage{graphicx}
\usepackage[usenames]{color}
\usepackage{multirow}
\usepackage{verbatim}
\usepackage{amsmath}
\usepackage{amssymb}
\usepackage{mathrsfs}
\usepackage{url}
\usepackage{float} 

\begin{document}

\title{Three Dimensional Metallic and Two Dimensional Insulating Behavior in Tantalum Dichalcogenides}

\author{Pierre~Darancet}
\email{ptd2105@columbia.edu}
\affiliation{Department of Applied Physics and Applied Mathematics, Columbia University, New York, NY 10027, USA}
\affiliation{Department of Physics, Columbia University, New York, New York 10027, USA}

\author{Andrew~J.~Millis}
\email{millis@phys.columbia.edu}
\affiliation{Department of Physics, Columbia University, New York, New York 10027, USA}

\author{Chris~A.~Marianetti}
\email{cam2231@columbia.eduu}
\affiliation{Department of Applied Physics and Applied Mathematics, Columbia University, New York, NY 10027, USA}

\date{\today}

\begin{abstract}
Using density functional theory with added on-site interactions (DFT+U), we study the electronic structure of bulk, monolayer, and bilayer of the layered transition-metal dichalcogenide $1T-TaS_2$. We show that a two-dimensional spin--$\frac{1}{2}$ Mott-phase exists for the monolayer in the charge-density wave state (CDW)  and that such a phase is systematically destroyed by packing of the distorted layers leading to a one dimensional metal for bulk, CDW-distorted TaS$_2$. The latter finding is in contrast with previous DMFT predictions --disagreement  we explain  by the weak effective interaction felt by the electrons in the CDW state. Experimental observations of insulating behavior may arise from disorder due to stacking faults. 
\end{abstract}
\maketitle

Layered transition metal dichalcogenides (TMDC) exhibit a wealth of competing phenomena, ranging from charge density waves (CDW)~\cite{DiSalvo,WilsonReview}, to metal-insulator transitions~\cite{WilsonReview,FazekasTosatti}, and superconductivity~\cite{Sipos}. Recent progress in mechanical exfoliation and device fabrication now allows for electrical characterization and gating of planar samples as thin as one unit cell~\cite{NovoselovPNAS}, opening new avenues for the study of basic physics and the  integration TMDC materials into functional components  of planar devices.  An appealing  property of  TMDCs  with respect to other two-dimensional materials such as graphene is their  variable bandgap. Some TMDC compounds have optical absorption spectra well matched to the solar spectrum and in the TMDC material molybdenum disulfide, the  passage from bulk material to few layer compounds  has  been experimentally shown to impact the magnitude and the nature of the gap ~\cite{Heinz}. Considering the TMDC family as a whole, the multiplicity of competing energy scales including  interlayer coupling and electron-electron and electron-phonon interactions suggest a high degree of tunability of the optoelectronic properties which requires investigation. 

Among TMDCs, $1T-TaS_2$ and $1T-TaSe_2$ are of particular interest for their interplay of charge density wave (CDW) and Mott physics. Below  a critical temperature of $T_C$ (180K for $1T-TaS_2$~\cite{Sipos} and 350K for $1T-TaSe_2$~\cite{ColonnaPRL}) the materials exhibit a so-called ``Star-of-David'' CDW involving an in-plane, $\sqrt{13}\times\sqrt{13}R=12.4^{\circ}$ periodic-lattice distortion (PLD). For $1T-TaS_2$, this PLD coincides with an increase of the resistivity \cite{Sipos}
and the appearance of an in-plane gap \cite{Dardel,PerfettiPRL1,PerfettiPRL2,HellmanPRL,KimPRL,StoltzPRB,ColonnaPRL}. This behavior has generally been attributed to the opening of a correlation (Mott) gap on  the $Ta-d_{z^2}$ subband localized at the star centers in the CDW state~\cite{FazekasTosatti,PerfettiPRL1,PerfettiPRL2,PerfettiNJP}, although some works have instead attributed it to a transition from a 2D, in-plane metal to 1D, out-of-plane metal with an Anderson-type transition due packing disorder of the centers of the distortions \cite{DiSalvoGraebner,DardelPRB}. 

In this letter, we use density functional (DFT) and density functional theory with added on-site interactions (DFT+U)
 methods to contrast single-layer, bilayer and bulk  $1T-TaS_2$  in its  CDW state.   We find that  bulk $1T-TaS_2$ is an out-of-plane metal rather than a Mott insulator, in disagreement with the conventional wisdom \cite{PerfettiPRL1,PerfettiPRL2,PerfettiNJP} but in agreement with some previous interpretations of experimental data~\cite{DardelPRB,DiSalvoGraebner}, while the monolayer is a Mott insulator with negligible magnetic exchange interactions and the bilayer compound  exhibits a dimer singlet phase. The key to the physics is the Star of David distortion: each unit cell of the distorted structure hosts one orbital at the Fermi level. While the frontier orbital in one cell is localized to the point that hybridization the frontier orbitals in adjacent cells in the same plane is negligible, the orbital is sufficiently delocalized that (as we show in detail below) the effective Coulomb interaction is remarkably weak, in particular smaller than the interplane bandwidth.  Our work shows how the interplay of dimensionality, correlation strength, and lattice distortion can affect bandgaps and magnetic properties in this important class of materials. 

\begin{figure*}[t]
	\begin{center}
	\includegraphics[width=0.90\textwidth]{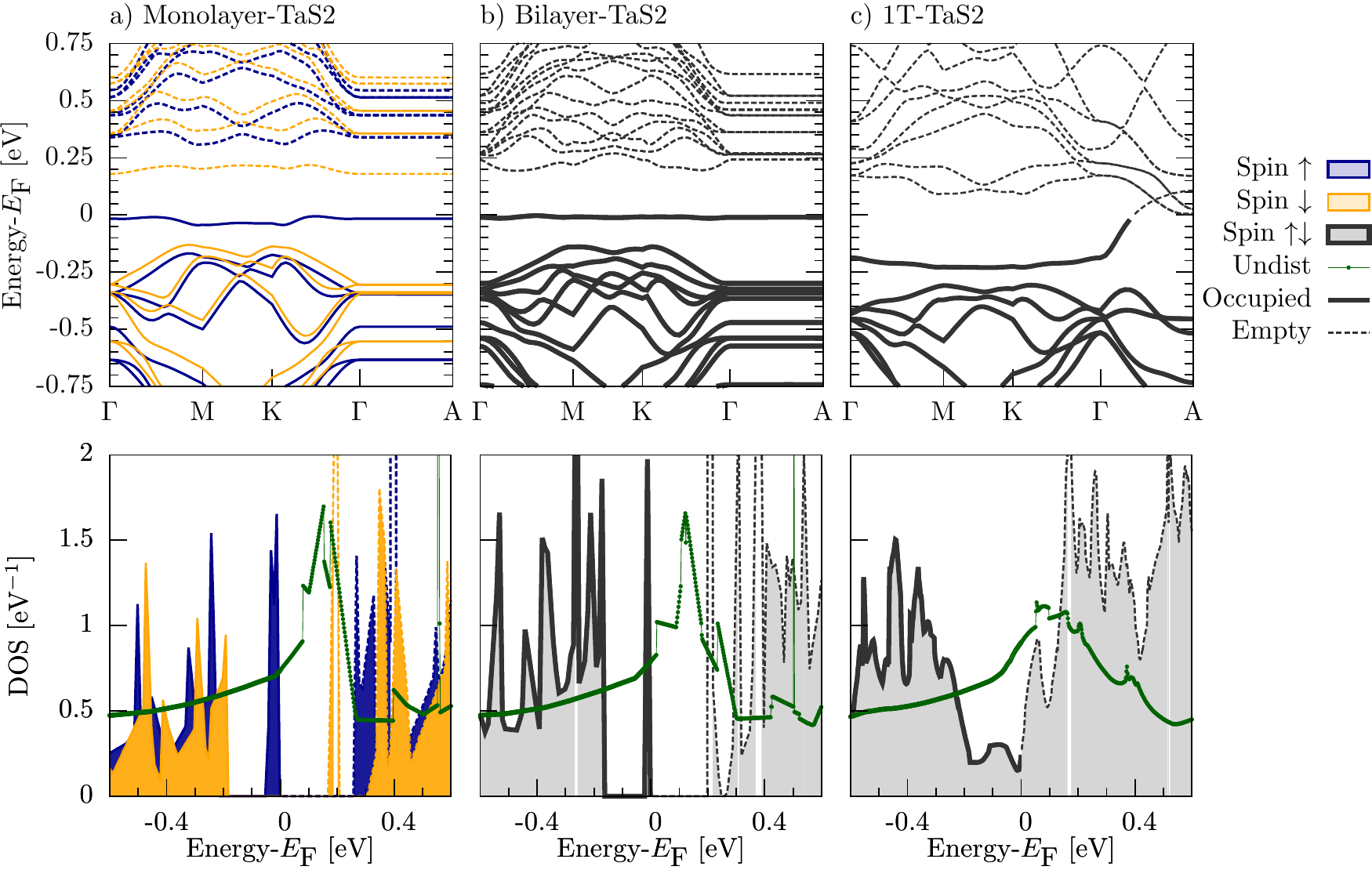}
	\end{center}
	\caption{Spin-resolved bandstructures (top) and density of states (bottom) for the distorted Monolayer (a), Bilayer (b), and Bulk (c) of $TaS_2$ computed with GGA+U (U=2.27eV) for vertically-stacked centers of distortion. The density of states for the undistorted structures are indicated with a dotted green line.}
	\label{Fig1:BS} 
\end{figure*}

The DFT calculations presented here were performed using the Vienna ab-initio simulation package (VASP) \cite{VASP1,VASP2,VASP3,VASP4}, with a GGA+U \cite{GGAU} approach and a plane-wave basis and projector-augmented wave (PAW) potentials \cite{PAW}. We used an energy cutoff of 350.0 eV, the tetrahedron method with a $14\times14\times8$ k-point mesh with respect to the primitive cell for the computation of the self-consistent electronic densities and of the structure relaxations, and with a denser, $28\times28\times16$ mesh for the computation of the densities of states.  To more accurately represent electronic correlations, an on-site U was included for the tantalum $5d$ orbitals. The value of $U=2.27eV$ was calculated using a linear-response method \cite{ULinearResponse} for the undistorted bulk. For all calculations, we relaxed the unit cell in  the in-plane direction while keeping the unit cell constant along the c-axis. The Star of David phase of bulk $1T-TaS_2$ was modeled using a $\sqrt{13}\times\sqrt{13}\times2$ unit cell containing 26 Tantalum atoms. Monolayers and bilayers were modeled using unit cells with the equivalent of 4 layers ($23.6\AA$) of vacuum along the c-axis to prevent unphysical interactions between periodic images. For all the systems, the PLD ground-states were found by relaxing the atomic positions from a randomized  version of their high symmetry positions in the $\sqrt{13}\times\sqrt{13}$ unit cell. 

In agreement with previous studies of bulk  $1T-TaS_2$ \cite{prb_79_220515_2009_liu,prb_82_155133_2010_liu} at $U=0$, the Star of David structure is found to be more stable than the undistorted structure by 11meV/$TaS_2$. The Star of David structure is  found to be more stable than the undistorted structured for all values of the on-site electron-electron interaction and interlayer distance considered in this paper and  is also obtained for non spin-polarized calculations. The energy gain on distortion is almost the same for the spin-polarized and non spin-polarized calculations (difference $<0.1$meV/$TaS_2$) suggesting that the distortion is not due to correlation effects in the Ta d-shell. We find a similar  Star of David structure  in the the monolayer and bilayer of $TaS_2$, where the Star of David distorted structures are found to be 19meV/$TaS_2$ and  20meV/$TaS_2$ more stable respectively than their high symmetry counterparts. 

In figure~\ref{Fig1:BS}, we show the DFT+U-computed band-structures and density of states for the relaxed Star of David-distorted monolayer, bilayer, and bulk $TaS_2$ along with the density of states of the corresponding undistorted compounds. We see that the distortion affects primarily the states within approximately $\pm \approx 0.5eV$ of the Fermi level, opening a gap of $\sim 0.35eV$ in agreement with photoemission data ~\cite{Dardel,PerfettiPRL1,PerfettiPRL2,HellmanPRL} but leaving a narrow band of states near the Fermi level. The band of in-gap states arises from Ta $d_{z^2}$ orbitals and has a very weak in-plane ($\Gamma-M-K$) dispersion but in the bulk material disperses very significantly along the interplane direction, with  a c-axis bandwidth of $\approx 0.45eV$ and  a Fermi surface  crossing between $\Gamma$ and $A$.  Thus we conclude that the distorted phase of the bulk $1T-TaS_2$ is a one-dimensional metal. Previous single-band Hubbard model analyses using values $U$ comparable to the $2.27eV$ used here found Mott insulating behavior~\cite{PerfettiPRL1,PerfettiPRL2,PerfettiNJP};  we will explain the difference below.   

The metallic behavior is solely a consequence of the interplane dispersion: the bilayer and monolayer compounds are insulators. In the DFT+U approximation used here the monolayer is found to be a ferromagnet with a spin splitting of $\approx 0.18eV$, but the in-plane magnetic coupling is negligible and we interpret the result as indicating that monolayer $1T-TaS_2$ is a Mott insulator with Mott gap $\sim 0.2eV$. The bilayer compound is  antiferromagnetic, with opposite spin alignment on the two planes. The out-of-plane antiferromagnetic behavior is an artifact of the DFT+U approximation, which does not treat spin rotation invariance correctly, and we interpret that result as indicating that the spins in the two planes form a singlet state.   It is worth noticing that the correlation gap of  the monolayer (0.18eV), which one may identify with the effective $U$ of the Hubbard-like model describing the low energy physics, is found to be much smaller than on-site $Ta-d$ calculated value of  $U=2.27eV$. 

\begin{figure}
	\begin{center}
	\includegraphics[width=0.49\textwidth]{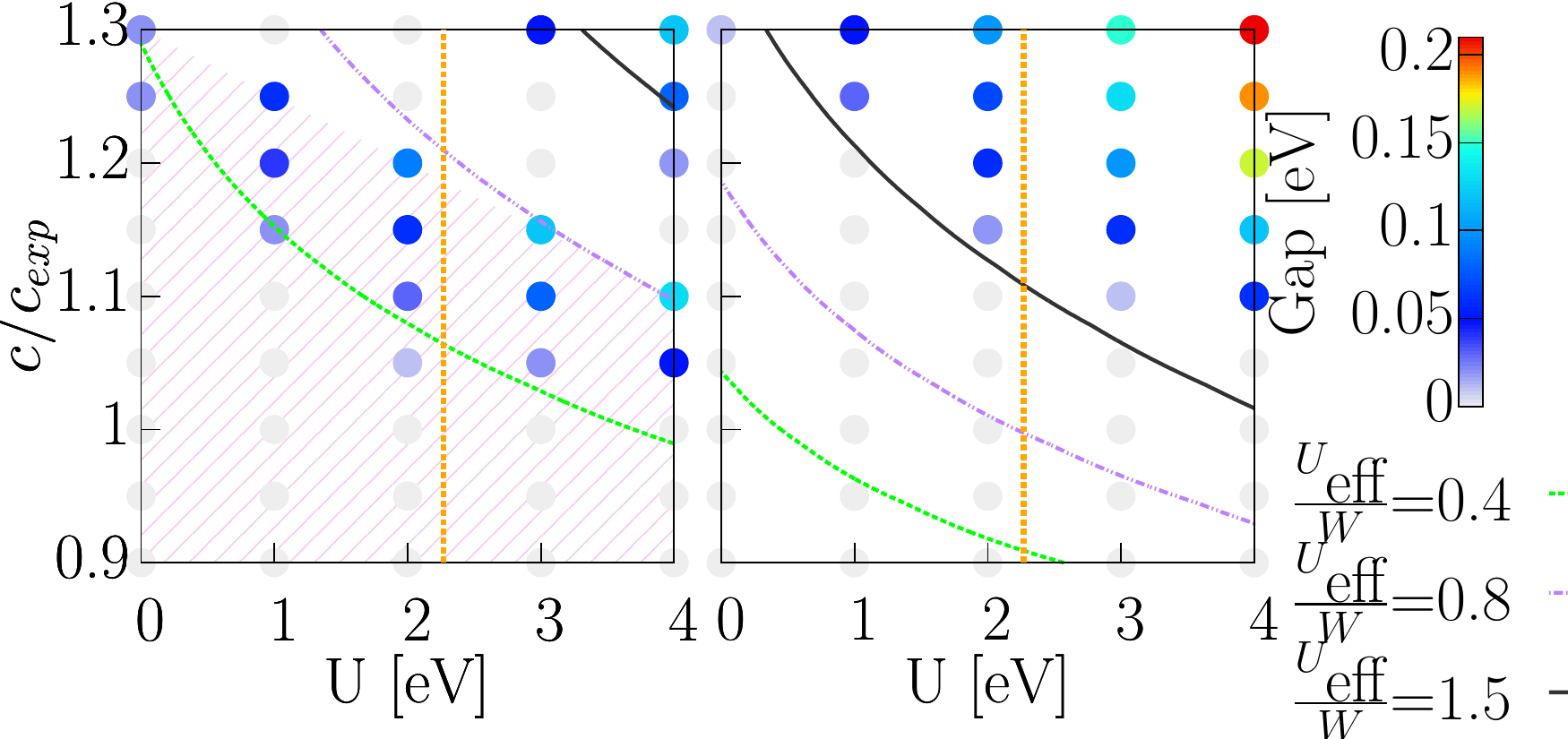}
	\end{center}
	\caption{Phase diagram of distorted  $1T-TaS_2$ in plane of atomic on-site electron-electron interaction U and interlayer distance c, for vertical (left) and trigonal (right) stackings of the centers of distortion. The contours in the left and right panels indicate the 0.4 (dotted, green on line), 0.8 (dashed, red on line) and 1.5 (solid, black on line) contours  of the parametrized $\frac{U_{\textrm{eff}}}{W}$ ratio ($U_{\textrm{eff}}$ is the effective interaction as defined in Eq.~\ref{Ueff}, $W$ is the out-of-plane bandwidth calculated in DFT for $U=0$). Gaps in the insulating state are indicated by colors. In each panel the vertical orange line indicates the calculated value U=2.27eV for undistorted bulk $1T-TaS_2$. The hashed and plain areas respectively indicate  out-of-plane antiferromagnetic and  ferromagnetic groundstates.}
	\label{PhaseDiagram} 
\end{figure}

To quantify the effect of the interlayer interactions and their competition with the electron-electron interactions, we show in Figure \ref{PhaseDiagram} the magnetic and metal-insulator phase diagram of bulk gap $1T-TaS_2$ as given in GGA+U   as a function of the onsite U and interlayer distance c, measured relative to the experimental value $c_{exp}=5.897\AA$. We consider  two choices of interplane stacking of the CDW distortion: a vertical stacking in which the centers of the Stars of David line up from layer to layer and a trigonal stacking ($Ta-I$ upon $Ta-III$) in which the centers are displaced.  At large c values ($c > 1.3c_{exp}$) the system becomes essentially identical to the monolayer:  a Mott insulator with a gap of $\simeq 0.08\times U$ [eV].  For the trigonal stacking (right panel) the system is ferromagnetic for all parameter values considered.  An apparently second-order phase transition (solid line, black on line,  $U_{eff}/W=1.5$) separates a Mott insulator from a ferromagnetic metal. For the vertical stacking (left panel) the phase diagram is more complicated, with the large U, large c   Mott insulator undergoing a transition to a reentrant paramagnetic metal phase (phase boundary approximately coincides with solid line, black on line, $U_{eff}/W=1.5$) which is separated from the small U antiferromagnetic metal phase  by an intermediate antiferromagnetic insulating  phase approximately bounded by the dashed and dotted lines ($U_{eff}/W=0.8$ and $U_{eff}/W=0.4$).

At the theoretically obtained atomic intra-d $U$ value $\approx 2.27eV$ (vertical lines) the metal-insulator transition occurs at $c\approx 1.15c_{exp}$ for the trigonal stacking and $c\approx 1.05c_{exp}$ for the vertical stacking. We suggest that pressure (to decrease the lattice constant) and intercalation ~\cite{Intercalation} (to increase it) experiments would be very interesting.

\begin{figure}
	\begin{center}
	\includegraphics[width=0.47\textwidth]{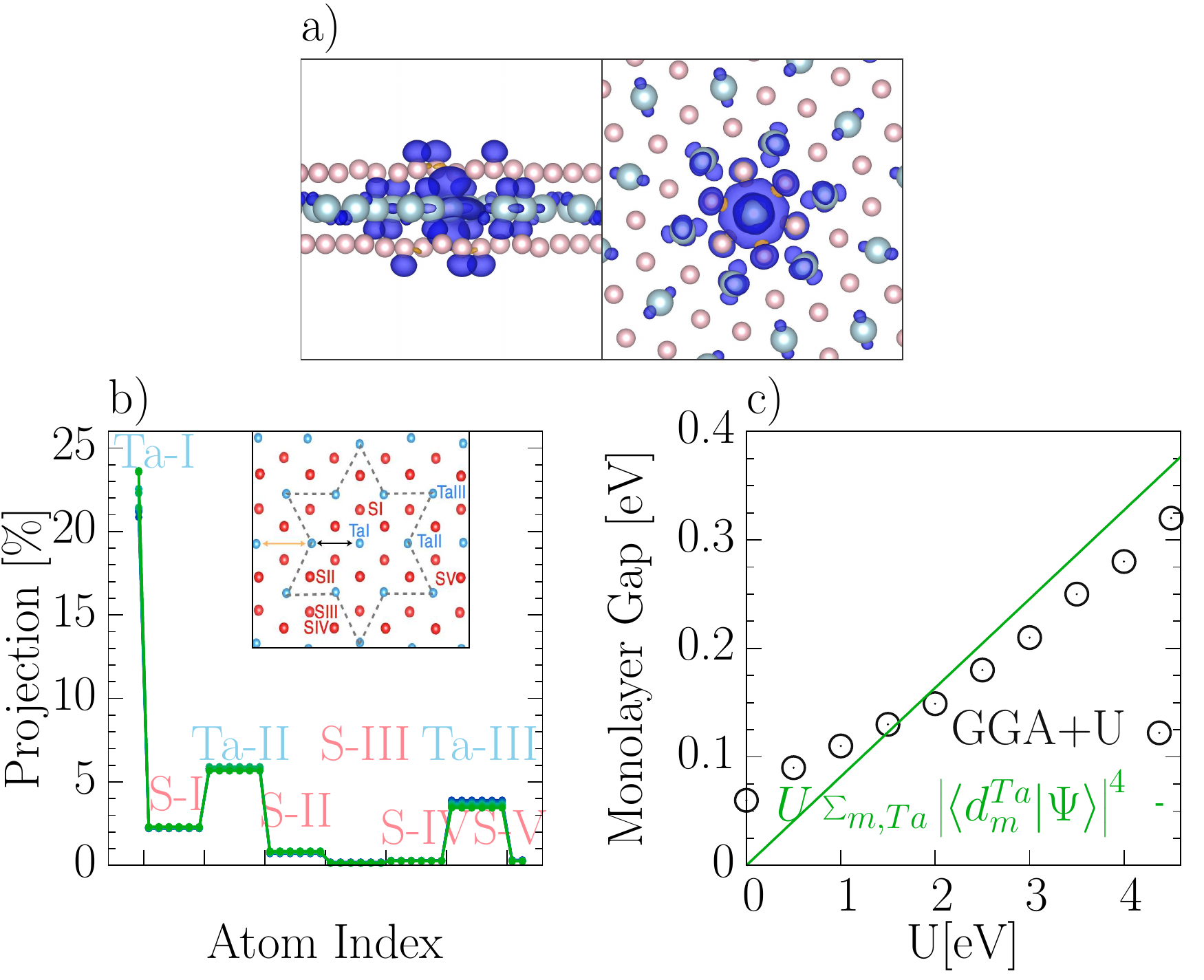}
	\end{center}
	\caption{ $a)$ Side- and top-views of the  spin-density for monolayer $TaS_2$. Spin up  (down) isosurfaces are  indicated in blue (orange); $b)$ atom-projected integrated charge density in the range $\left[ E_F-0.1 ; E_F\right]$ for different values of the on-site electron-electron interaction U ($U \in \left[ 0eV ; 4eV\right]$), for the monolayer of $TaS_2$.  $c)$ Corresponding effective electron-electron interaction $U_{\textrm{eff}}$ compared with the gap of the monolayer computed in DFT. }
	\label{Fig2:WF} 
\end{figure}

To understand these findings we present in the top panel of Fig.~\ref{Fig2:WF} the spin density of the monolayer system. Because in the DFT+U approximation used here the monolayer is a fully polarized ferromagnet, the spin-density is equivalent to the charge density associated with the lower Hubbard band. 
The spin-density is centered around the center of the Star of David, but with a non-negligible weight on the neighboring tantalum and sulfur atoms, and significant spreading along the out-of-plane direction. The atomic projection of the lower-Hubbard band of the monolayer, obtained by integrating the atom-projected density of states in the range $\left[ E_F -0.1 : E_F \right]$ is presented  in Figure \ref{Fig2:WF}, b). Though 79\% of the lower-Hubbard band lies on the tantalum atoms, only 20 to 25\% is localized at the center of the distortion ($Ta-I$, following the notation in Figure \ref{Fig2:WF}, c), inset), while other twelve tantalum atoms each have non-negligible projections around 4-6\% (these values are found to be weakly dependent on U).

This in-plane delocalization is the origin of the weak effective electron-electron interaction ($0.18eV$) relevant to the band in the gap. In essence, the small amplitude for the electron to be localized on any given Ta site of the Star of David implies that the effective interaction $U_{\textrm{eff}}$ is much smaller than the basic on-site Ta interaction $U$. The difference  may be quantified by projecting the Wannier  function $|\Psi>$ of the band in the gap onto the local d-orbitals $|d_{am}>$ (here $a$ labels the Ta sites in one unit cell and $m$ is the angular momentum quantum number)
\begin{equation}
\frac{U_{\textrm{eff}}}{U}=\sum_{a\in \star,m}|<d_{am}|\Psi>|^4
\label{Ueff}
\end{equation} 
As seen in Figure~\ref{Fig2:WF}, c), evaluation of Eq.~\ref{Ueff} leads to an effective interaction much smaller ($\sim 8\%$) than the on-site U, in good agreement with the gaps of the monolayer obtained for  GGA+U calculations at different U values (The deviations visible at very small U values arise from the correlations already present in GGA). For the calculated atomic value of $U\approx 2.27eV$, this effective electron-electron interaction $U_{eff}=0.18eV$ is larger than the in-plane bandwidth (70meV), explaining the Mott insulating nature of behavior of the monolayer, but is much weaker than the out-of-plane bandwidth of the bulk (0.45eV), indicating that the bulk material should not be considered as a Mott insulator. These results suggest that the reported insulating behavior of bulk TaS$_2$ arises from disorder  in a one dimensional conduction band as previously proposed ~\cite{DardelPRB,DiSalvoGraebner}. In this regard, we remark that the interplane hopping in the trigonally stacked structure is a factor of $\sim 3$ smaller than in the vertically stacked structure, suggesting that the localizing disorder may originate from stacking faults.

The interplay between hopping and interaction leads to interesting results in the bilayer case, where two vertically displaced Star of David units may be modeled as a two-site Hubbard model, with interaction $U_{\textrm{eff}}$ and hopping $t$ equal to one quarter of the c-axis bandwidth, i.e.   $t\approx 0.1eV$ for the vertically stacked case and $\approx 0.035eV$ for the trigonally stacked case. Solving the resulting two site Hubbard model leads to an even-parity singlet ground state, with a triplet excited state about $0.08eV$ (vertical) and $0.012eV$ (trigonal) higher, and an optical gap (relevant for E-fields applied perpendicular to the plane) of $0.25eV$ (vertical) or $0.19eV$ (trigonal).    The sensitivity of the gaps to the interlayer hopping amplitude suggests that optical and magnetic  transitions in the bilayer compound may be tuned by intercalation, pressure, and stacking, and potentially leading to an interesting set of excitonic transitions.

By approximating the bandwidth $W$ by the bandwidth of the non-spinpolarized distorted bulk at U=0, and  $U_{\textrm{eff}}$ as the gap of the monolayer, we find that GGA+U gives the metal-insulator transition for the ferromagnetic ground state  at $\frac{U_{\textrm{eff}}}{W}\simeq 1.5$,  in good agreement with the critical value $\frac{U_{\textrm{eff}}}{W}\simeq  1.3$ for the metal-insulator transition emerging from DMFT calculations  \cite{PerfettiPRL1,PerfettiNJP}. We therefore suggest that the disagreement between our results and those of Refs.  \cite{PerfettiPRL1,PerfettiNJP} arises in part from an excessively large value of U assumed in those references. Moreover the sensitivity of the location of  the metal-insulator phase boundary to the nature (ferromagnetic vs antiferromagnetic) of the magnetic state suggests that any insulating states that do occur in the $1T-TaS_2$ family of materials be regarded as arising more from out-of-plane antiferromagnetic order than from the Mott phenomenon per se. Moreover, we observe that the dependence of magnetic ordering on stacking of distortions, as well as the small effective $U$ values, suggests that mappings onto Hubbard models be regarded with caution. The details of the underlying wave functions and of nearby perhaps virtually occupied states, which are not easily represented in a Hubbard model, will be important. Finally, an important challenge raised by our work is understanding the photo-induced dynamics of the out-of-plane metallic bulk $1T-TaS_2$, as the observed collapse of the in-plane gap happens on a timescale inconsistent with the dynamics of a Peierls insulator~\cite{PerfettiPRL2}.

In conclusion, we have shown that the electronic structure of $1T-TaS_2$ in its CDW state strongly depends on interlayer interactions. In particular, we found that, upon exfoliation from the bulk, the distorted monolayer of $TaS_2$ undergoes a metal-insulator transition, associated with electrons localizing in in-plane 13-atoms clusters.  Finally, we have revisited the nature of the bulk  $1T-TaS_2$, that we predict to be a band-insulator in plane and metallic out-of-plane upon distortion, and explained the disagreement with previous interpretations by the weak effective electron-electron interactions felt by the electrons delocalized across the Star of David. The monolayer compounds are predicted to be Mott insulators with a $s=1/2$ degree of freedom in each unit cell of the CDW structure, while the bilayers form a singlet state with a tunable optical gap. 

We thank Prof. Abhay Pasupathy and Prof. James Hone for fruitful discussions. This work was funded by NSF under contract DMR-1122594 and used resources at the New York Center for Computational Sciences at Stony Brook University/Brookhaven National Laboratory which is supported by the U.S. Department of Energy under Contract No. DE-AC02-98CH10886 and by the State of New York.



\begin{thebibliography}{99}




\bibitem{DiSalvo}
JA Wilson, FJ Di Salvo, S Mahajan, 
Advances in Physics, 1975 - Taylor \& Francis.


\bibitem{WilsonReview}
J.A. Wilson, A.D. Yoffe,
Advances in Physics  \textbf{18,}  73, (1969).

\bibitem{FazekasTosatti}
P. Fazekas, E. Tosatti 
Philosophical Magazine Part B, \textbf{39,}  3  (1979);
P. Fazekas, E. Tosatti, 
Physica B+C, \textbf{99,}  14,  183-187 (1980).

\bibitem{Sipos}
B. Sipos et al., 
Nature Materials, \textbf{7,} 960 (2008). 


\bibitem{NovoselovPNAS}
K. S. Novoselov et al., PNAS \textbf{102,} 10451 (2005).



\bibitem{Heinz}
F.K. Mak et al., 
Physical Review Letters \textbf{105,} 13 (2010): 136805.


\bibitem{Dardel}
B. Dardel et al., Physical Review B \textbf{45,} 1462 (1992);






\bibitem{HellmanPRL}
S. Hellmann et al., 
Physical Review Letters \textbf{105,} 187401 2010 

\bibitem{PerfettiPRL1}
L. Perfetti et al.
Physical Review Letters \textbf{90,} 166401 (2003)


\bibitem{PerfettiPRL2}
L. Perfetti et al.
Physical Review Letters \textbf{97,} 6 (2006).


\bibitem{KimPRL}
J.J. Kim et al., Physical review letters \textbf{73,} 15 (1994).

\bibitem{StoltzPRB}
D. Stoltz et al., 
Physical Review B \textbf{76,} 7 (2007).

\bibitem{ColonnaPRL}
S. Colona et al., 
Physical Review Letters \textbf{94,}  3, 036405 (2005).


\bibitem{PerfettiNJP}
L. Perfetti et al. 
New Journal of Physics \textbf{10,}  5 (2008). 


\bibitem{DardelPRB}
 B. Dardel, et al.
 Physical Review B \textbf{45,}  3 1462 (1992).

\bibitem{DiSalvoGraebner}
F. J. Di Salvo, J. E. Graebner
Solid State Communications \textrm{23,} 11  825-828 (1977).


\bibitem{VASP1}
 G. Kresse, J. Hafner, Physical Review B \textbf{47,} 558 (1993).
\bibitem{VASP2}
G. Kresse, J. Hafner, Physical Review B \textbf{49,} 14251 (1994).
\bibitem{VASP3}
G. Kresse, J. Furthmuller, Computational Materials Science \textbf{6,} 15 (1996).
\bibitem{VASP4}
G. Kresse, J. Furthmuller, Physical Review B \textbf{54,} 11169 (1996).

\bibitem{GGAU}
S. L. Dudarev, et al., 
Physical Review B \textbf{57,} 1505 (1998).



\bibitem{PAW}
G. Kresse, D. Joubert, Physical Review B \textbf{59,} 1758 (1999).


\bibitem{ULinearResponse} 
M. Cococcioni, S. de Gironcoli, Physical Review B \textbf{71,} 035105 (2005) 


\bibitem{prb_82_155133_2010_liu}
G. Yizhi, A. Y. Liu, Physical Review B \textbf{82,} 155133 (2010) 


\bibitem{prb_79_220515_2009_liu}
A. Y. Liu, Physical Review B \textbf{79,} 220515 (2009) 




\bibitem{Liechtenstein} 
 A. I. Liechtenstein, V. I. Anisimov, and J. Zaane, Physical Review B \textbf{52,} R5467 (1995).


\bibitem{Intercalation}
H. I. Starnberg,  
Modern Physics Letters B \textbf{14,} 3, 455-471 (2000).









\end{thebibliography}

\end{document}